\documentclass[twocolumn]{svjour3}
\usepackage{shadethm}

\usepackage{savesym}
\usepackage{amsmath}
\savesymbol{iint}
\usepackage{esint}
\restoresymbol{TXF}{iint}

\usepackage{latexsym}
\usepackage{amsmath}
\usepackage{amssymb}
\usepackage{graphicx}
\usepackage[pdftex,
            pagebackref=true,
            colorlinks=true,
            linkcolor=blue
           ]{hyperref}
\usepackage{float}
\usepackage{alltt}
\usepackage{authblk}

\begin{document}
\title{Fractal social organization as a foundation to\\pervasive social computing services}
\author{Vincenzo De Florio}
\institute{Global Brain Institute and Evolution, Complexity and COgnition research group\\
       Vrije Universiteit Brussel, Belgium\\
       \email{vincenzo.deflorio@gmail.com}}
\date{} 
\maketitle %

\begin{abstract}
Pervasive social computing is a promising approach that promises to empower
both the individual and the whole and thus candidates itself as a foundation
to the ``smarter'' social organizations that our new turbulent and
resource-scarce worlds so urgently requires.
In this contribution we first identify
those that we consider as the major requirements to be fulfilled in order
to realize an effective pervasive social computing environment.
We then
conjecture that our service-oriented community 
and fractal social organization fulfill those
requirements and therefore constitute an effective strategy to design pervasive social computing
environments.
In order to motivate our conjecture, in this paper
we discuss a model of social translucence and discuss fractal social
organization as a referral service empowering a social system's parts and whole.
\end{abstract}

\section{Introduction}
Several are the definitions of
pervasive social computing (PSC) that may be found in the literature.
Those definitions provide different pictures of what PSC is and which challenges it is meant
to tackle. A key difference is that
some definitions propose PSC as an approach to augment the individual while
others put the accent on the social dimension. A related major difference 
is given by the
challenges that PSC is to tackle.
Those challenges range from the creation of an
``an integrated
computing environment, which promises to augment five facets of human intelligence:
physical environment awareness, behavior awareness, community awareness, interaction awareness, and content awareness''
to the definition of an approach
``born for addressing new situations and new challenges in the age of integrated cyber and physical worlds''.
Our modern times have indeed introduced several new situations that challenge our ability to sustain
our societies and economies~\cite{DeBl08a}. Those situations and challenges may be synthetically expressed as
the advent of Anthropocene---the time of man as the pivotal element
for the progress and the evolution of our ecosystems. In the times of Anthropocene, it is man the main factor
who is responsible for the emergence of ecosystem stability or instability; it is man's action that can lead
either to balance and sustainability or to rapid resource exhaustion, chaos, and unsustainability.
The ever growing human population; a physical world more and more depleted of its resources; and
the widespread of the pure individual-centric, competitive-oriented model,
all themes masterfully discussed by Hardin in his ``Tragedy of the Commons''~\cite{Hardin68},
provide us with a pessimistic vision to our future. The negative role of man in this context were
brilliantly rendered by famous cartoonist and comic artist Walt Kelly
for the 1970 edition of Earth Day, in which his character Pogo after observing the
devastating effect of pollution on their habitat concludes: ``We have met the enemy, and he is \emph{us}!''

\begin{figure*}[t]
\centerline{\includegraphics[width=1.0\textwidth]{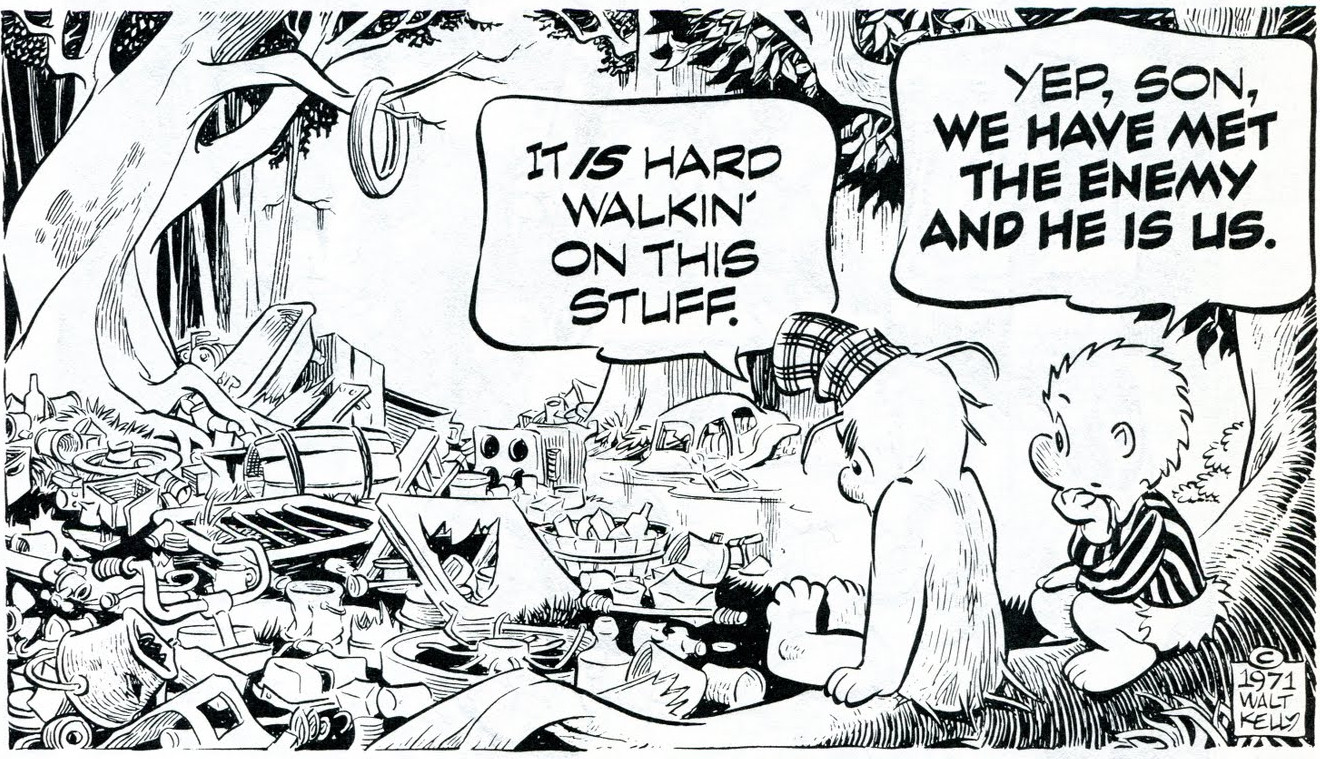}}
\caption{To be added.
Copyright Okefenokee Glee \& Perloo, Inc.  Used by permission.%
}
\end{figure*}

I believe such a pessimistic conclusion should be coupled with other facts of opposite sign.
Our modern times are providing us with unprecedented possibilities for more profitable and sustainable forms
of social action. As observed by~\cite{BenCa09},
the ``pervasiveness of handheld devices and the enormous popularity of social networking websites''
set the conditions to the emergence of novel, more efficient and cost-effective ways to organize our services.
Moreover, a technology ``relentlessly pushing down communication costs''~\cite{Kellogg05} provides us with the
opportunity to evolve our ``static, local organization of an obsolete yesterday''~\cite{BuckFuller75} and
realize and experiment with novel, dynamic, and distributed forms of organization.
Our stance in the present contribution is that PSC may be interpreted also as
the technological foundation for the definition of the above-mentioned novel, dynamic, and distributed
forms of organization.

In fact, PSC artificially augments several ``facets of human intelligence''~\cite{Zhou12}
and realizes a quasi-ubiquitous ``communication channel'' that, in the classic definition
by Kenneth Boulding~\cite{Bou56}, turns a set of roles into a
\emph{social organization},
namely the most advanced known system class in his general systems classification.

This paper discusses fractal social organization (FSO) as an architecture for the design of
PSC services. We first discuss social
organizations in the framework of PSC in Sect.~\ref{s:socorg}.
FSO is then introduced in Sect.~\ref{s:socfso}.
Next, in Sect.~\ref{s:found}, we discuss the relation between FSO and PSC.
Our conclusions and a view to are finally given in Sect.~\ref{s:end}.

\section{Social organizations}\label{s:socorg}
One of the first and most renowned definitions of social system was provided in 1956 by Kenneth Boulding
in his now classic article~\cite{Bou56}:
\begin{quote}
``it is tempting to define social organizations, or almost any social system,
  as a set of roles tied together with channels of communication.''
\end{quote}
After 70 years from its formulation, Boulding's definition is still actual and deserving analysis.
The definition includes four parts:
\begin{description}
\item[\textbf{A set:}] This highlights the fact that social actors are usually grouped
into sets, defined by some measure of physical or logical proximity.
In fact this is a subset of those social actors that are relatively close to each other
according to the above measure. The term \emph{locality\/} has been used to refer
to the locus defined by the physical or logical measure of proximity~\cite{BenCa09}.
In the case of a \emph{pervasive\/} social system this implies that this set is relatively dense
and populated.
\item[\textbf{of roles:}] Quoting again from~\cite{Bou56},
\begin{quote}
``The unit of such systems is not perhaps the person---the individual human as such---but the `role'---that
  part of the person which is concerned with the organization or situation in question.''
\end{quote}
Shifting the attention from the actor to the role introduces another source of dynamic behavior:
a social system is one in which roles may mutate with time and the context, leading to complex
dynamical reconfigurations.
\item[\textbf{tied together:}] This part of the definition highlights the fact that in a social ``whole''
the parts are joined together by some aggregative force. This force may be due to cultural,
anthropological, or physiological reasons. Symbiotic or other mutualistic relationships; affection;
sharing of the same ideals or aims; lineage; and, in general, the individual returns produced by the social union
strenghten the cohesion of the parts with the social whole. A dual, disgregative, force exists,
which corresponds to the strength of the negative returns that are experienced by the social parts.
If the perceived disadvantages resulting from the social union outweight the perceived advantages,
the parts shall loosen up from the whole, disintegrating the social system into its constituent units.
Because of this, ``tying the roles together'' means also the ability to accentuate the ``centripetal''
social force and to dump the ``centrifugal'' force. A technique to achieve this is
\emph{social translucence}: a socially translucent object, service, system, or user,
is one for which the returns associated with social interactions are made apparent~\cite{ErKe00}.
\item[\textbf{with channels of communication:}] 
It is once again Boulding who observes how
\begin{quote}
``Communication and information processes are found in a wide variety of empirical situations,
  and are unquestionably essential in the development of organization, both in the biological and the social world.''
\end{quote}
Several aspects should be highlighted here.
First, ``communication'' is here much more than sharing data---it is sharing structured,
semantically described information~\cite{BenCa09} that provide snapshots of the individual and social context:
capabilities, policies, availabilities, location data, as well as viewpoints about experienced facts.
Secondly, nowadays powerful and pervasive ``communication channels'' are driven by a technology that offer ever more
complex services while ``relentlessly pushing down communication costs''~\cite{Kellogg05}.
Third, the ``pervasiveness of handheld devices and the enormous popularity of
social networking websites''~\cite{BenCa09}
contribute to the creation of
a quasi-ubiquitous and world-wide ``communication channel'' that
in practice artificially augments several ``facets of human intelligence''~\cite{Zhou12}.
\end{description}

It is the ambition of pervasive social computing (PSC) to realize such social system or better, using
Boulding's terminology, to realize such social organization. PSC provides a ``communication channel''
that energizes aggregative forces with the following extra advantages:
\begin{description}
\item[\textbf{Social translucence:}] thanks to the PSC environments, the users
can more easily ``see through'' and ``farther'', which allows them to identify opportunities
resulting from social unions~\cite{ErKe00,Kellogg05,de2014mutualistic}.
\item[\textbf{Referral Service:}]
As exemplified in~\cite{BenCa09}, through PSC
one does not look for roles; rather, one advertises one's task.
It is then the PSC environment that identifies actors best-matching the sought roles.
PSC thus shifts the responsibility for handling social services from the user to the ``channel''.
It is the channel that is charged with the task to manage the highly dynamic set of roles
made available by the actors.
%
%
\item[\textbf{Empowering the Parts:}]
PSC ``socially augments'' the individual with
``five facets of human intelligence:
  physical environment awareness, behavior awareness,
  community awareness, interaction awareness, and content awareness''~\cite{Zhou12}.
As a side-effect, this also strengthens the social ``whole'' because of the
extra advantages deriving from the augmented awareness.
\item[\textbf{Empowering the Whole:}]
PSC also empowers the union of the parts, making it possible to address
``new situations and new challenges in the age of integrated cyber and physical worlds''~\cite{Zhou12}.
More than this, PSC favors the emergence of a coherent and purposeful ``social behavior''
from a set of mostly independent individual behaviors.
A new ``social individual''---the social system---then raises from the union of the parts~\cite{DeBl10}.
The classic behavioral classification of~\cite{RWB43} or their extensions such as the one
introduced in~\cite{DF15b,DF15d} may then be used
to characterize the ``systemic class'' of the social system.
One of the challenges of PSC is to foster
the emergence of advanced forms of collective intelligence in PSC-empowered social systems.
\end{description}

Our stance here is that three are the key requirements to be fulfilled in order to realize
an effective PSC environment:
\begin{enumerate}
\item The definition of an effective ``communication channel'': a ``server'' taking the shape of,
e.g., a middleware~\cite{BenCa09} able to offer PSC services to the constituent actors of a
social system identified by a given locus.
\item The definition of an effective ``second-order'' communication channel able to
offer PSC services \emph{to a set of servers\/} as defined in 1.
\item The definition of a mechanism to ``tie together'' and effectively
organize a nested compositional hierarchy of communication channels.
\end{enumerate}

In what follows we conjecture that our service-oriented community (SoC)~\cite{DeBl10,DFCBD12}
and fractal social organization~\cite{DF13c,DF13a,DF14d,DF15c} fulfill the above three requirements
and therefore constitute an effective ``strategy'' to design PSC environments.
%
in next secton we introduce SoC's and FSO's in terms of the three above-mentioned PSC requirements.
%

\section{Service-oriented communities and fractal social organizations}\label{s:socfso}
We have concluded last section by identifying those that we consider to be the three key requirements to PSC.
In what follows we ``map'' those three requirements onto the concepts of SoC and FSO.

\subsection{Service-oriented communities as PSC channels}\label{s:fso:soc}
\begin{figure*}[t]
\centerline{\includegraphics[width=1.0\textwidth]{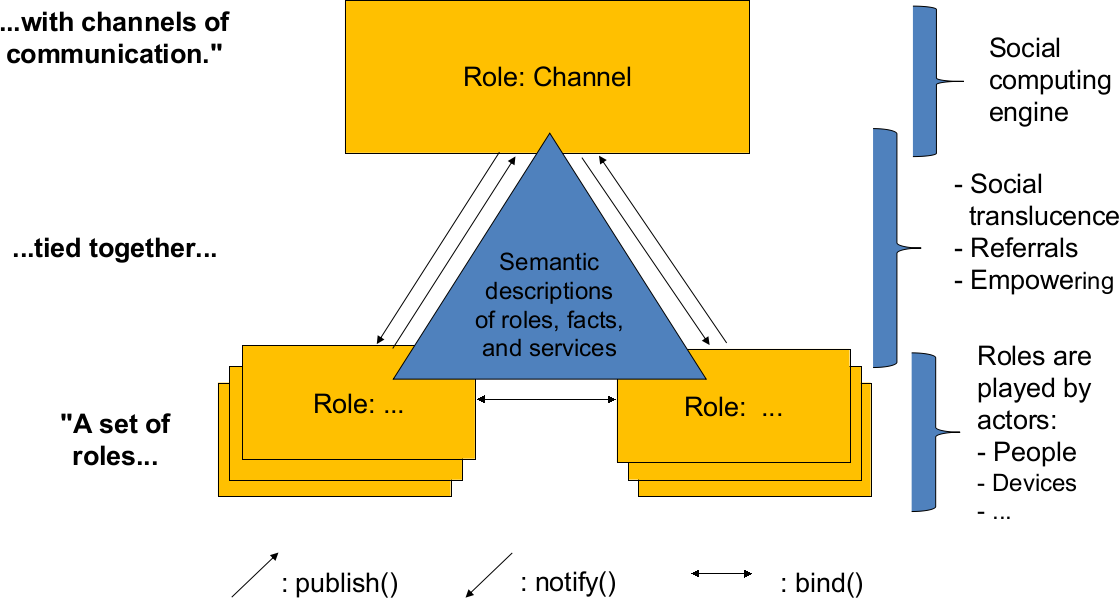}}
\caption{General structure of a service-oriented community. Rectangles represent actors and roles.
Actors/roles publish context information (roles, facts, and services)
via the \texttt{publish()} method.
The top rectangle
is the social computing engine, which manages the PSC services (translucence, referral, and empowering)
via semantic description and matching (represented as the blue triangle).
Candidate roles are informed via \texttt{notify()}. If the corresponding actors agree to commit to
a social service, they \texttt{bind()} together and the service is started.}\label{f:SoC-1}
\end{figure*}

A SoC may be described as the practical organization of a social system according to the classic
definition recalled in Sect.~1: ``a set of roles tied together via channels of communication''.
Figure~\ref{f:SoC-1} provides a high-level view to the structure of the SoC and highlights its relation
with social systems and PSC.
In a SoC, a set of roles played by human beings
and cyber-physical systems in physical or logical proximity are tied together via
a social computing engine (SCE).
By means of a publish/subscribe mechanism, said engine is made aware of: the dynamic assignments
of roles; the availability of the corresponding actors; their engagement policies; their location;
the occurrence of events; resource state changes; requests for service; and other contextual information.
The SCE receives said information as semantically described services, which are then stored into a
service registry. The arrival of new service descriptions triggers a semantic match with the records
already stored in the registry. This is done in order
to identify roles able to fulfill the requests for services
waiting to be answered. Once roles are identified, a notification is sent to the corresponding actors.
In so doing, \emph{social translucence\/} is realized:
actors become aware of the ``win-wins''---the mutually
rewarding relationships that they may enact and exploit.
At the same time, the SCE makes it possible to optimize
the tasks of the ``parts'' (the social actors) and those of the ``whole'' (the social system).
New arrivals are also matched against service role protocols, namely sequences of actions that are activated
by the verification of a guard expressing the availability of one or more roles. One such protocol, e.g.,
instructs what to do when an accelerometer ``fires'', publishing the fact that an associated elderly person
is suspected to have fallen~\cite{DeFPa15b}.
SCE also realizes a simple form of e-referral~\cite{EReferrals,ShBe07,Boudreau}:
instead of contacting possible actors ``on a one-to-one basis''~\cite{BenCa09},
requesting actors just ``advertise'' their requests. We refer to such service
as ``simple'' because the range of possible candidate roles is limited to those
in proximity of each other.

SoC's have been used in the past to realize cyber-physical societies such as
the mutual assistance community (MAC)~\cite{SDGB10b}. In a MAC, social translucence is
sought by identifying mutualistic relationships through the use of
semantic service description and matching~\cite{4618805}. One such relationship is the so-called
``participant'' service mode, which was exemplified in~\cite{SDGB07a} and modeled
in~\cite{de2014mutualistic}.

\subsection{Fractal social organization}\label{s:fso:fso}
FSO may be concisely defined as the organization of a nested compositional hierarchy (NCH) of SoC's.
In order to better explain this concept, we need to describe two elements: how the NCH is structured and how
it is organized.

\begin{figure*}[t]
\centerline{\includegraphics[width=1.0\textwidth]{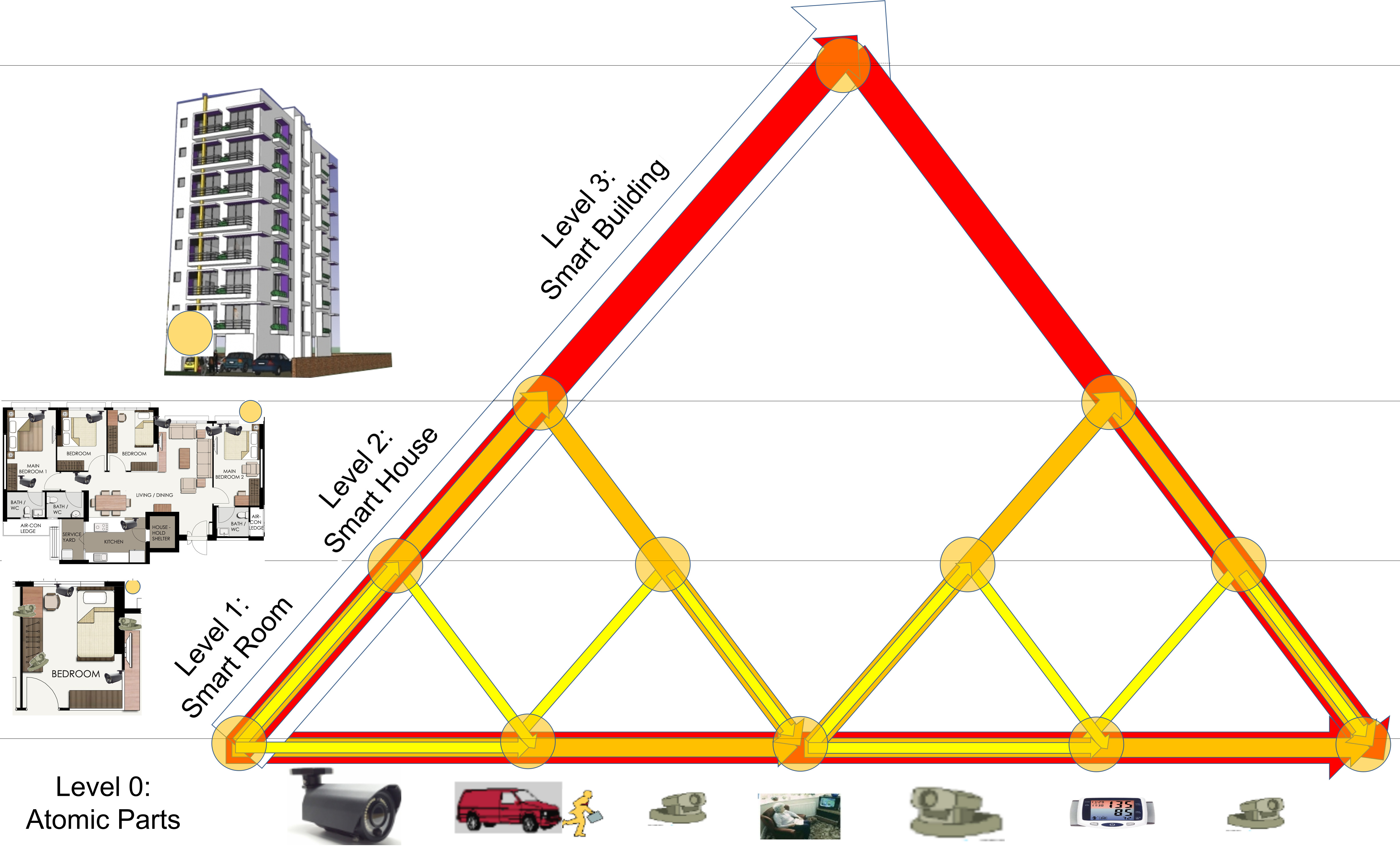}}
\caption{Exemplification of a 3-layer FSO. PSC loci are in this case smart rooms,
smart houses, and a ``smart building''. Top vertex represent SCE's, which manage their SoC
and at the same time are members of their parent SoC.}\label{f:FSO}
\end{figure*}

The NCH structuring is obtained by allowing an SoC to be a member of a greater SoC.
This means that sets of roles are included into greater sets of roles.
For simplicity, those sets may be visualized as loci, or containers,
structured as a Matryoshka doll.
The idea is exemplified in Fig.~\ref{f:FSO}. In that picture we have a set
of layers representing different classes of loci---for instance rooms, houses, and buildings.
The only difference between the loci lies in a simple compositional rule:
a building can contain houses, human beings and cyber-physical systems;
houses can contain rooms, human beings and cyber-physical systems; 
and rooms may only contain human beings and cyber-physical systems.
This may be generalized by stating that a level-$k$ SoC can only include 
level-$j$ SoC's, with $j$ and $k$ integers such that $0\le k<j$ and
the assumption that a level-0 SoC corresponds to an individual member---a human being or
a cyber-physical system.

The second element we need to clarify is the organization of the SoC hierarchy.
In order to do so, we now introduce a number of symbols and definitions.

\begin{definition}
Given any FSO $f$,
let us refer to
SoC${}_f(k)$ as to the set of all the $level-k$ SoC's of $f$. When $f$ may be omitted
without introducing ambiguity, we shall use symbol SoC$(k)$.
\end{definition}

\begin{definition}
Given any FSO $f$,
for any SoC $s\in f$, we shall refer to $\Pi(s)$ as to the parent of $s$, namely
the SoC of $f$ that includes $s$ among its members.
\end{definition}

\begin{definition}
Given any FSO $f$,
let us define $m_f = \max_k \hbox{SoC}_f(k)$. When $f$ may be omitted
without introducing ambiguity, we shall use symbol $n$
to refer to the root level of $f$.
\end{definition}

\begin{definition}
Let us define
as request for services the following guarded action:
\begin{equation}
a: r_1, r_2, \ldots, r_n,\label{e:ga}
\end{equation}
in which $r_1, r_2, \ldots, r_n$ are role identifiers and $n\ge0$ is an integer.
\end{definition}

\begin{definition}
Given any SoC $s$,
We shall say that request for service $a$ is enabled when it is in the service registry
of the SCE of $s$.
\end{definition}

\begin{definition}
Given any FSO $f$,
any SoC $s\in f$,
and any request for service $a$ that was enabled in $s$, we shall say that $a$ is active when
all of its role identifiers have been associated to actors of $f$.
\label{d:active}
\end{definition}

Note that, in Definition~\ref{d:active}, we refer to the actors of $f$, not $s$. This means that
those actors may belong to any of the SoC's of $f$.

\begin{definition}
An association of roles to actors shall be called in what follows as ``enrollment''.
A ``local enrollment'' shall be one in which all roles are found in the present SoC.
Otherwise, we shall use the term ``global enrollment.''
\end{definition}

As in sociocracy~\cite{BuEn12}, also in FSO we have two organizational rules:
\begin{description}
\item[Double membership:] For any $0<k<m$ and for any $s\in\hbox{SoC}(k)$, let us call $\sigma(s)$ the SCE of $s$.
The double membership rule states that $\sigma(s)$ is at the same time a member of $s$ and a member of $\Pi(s)$.
\item[Exception:] Being the SCE of $s$, $\sigma(s)$ receives notifications published by all the members of $s$.
When an incoming notification corresponds to a request for services $a$ defined as
in~\eqref{e:ga}, $\sigma(s)$ initiates a local enrollment, trying to identify a set of local actors
to be associated with the roles necessary to enable $a$.
This is done by checking the current state of the
system registry, which maintains an up-to-date state of all the members of $s$.
The exception rule states that, whenever $\sigma(s)$ cannot identify in $s$ all the roles necessary
to fulfill $a$, it will raise an \emph{exception\/} by publishing $a$ and its missing roles to $\Pi(s)$.
In other words, a failed local enrollment is turned into a global enrollment.
\end{description}

In what follows, we illustrate how the the SoC and the FSO fulfill the key requirements
to the realization of an effective PSC environment that we have
introduced in Sect.~\ref{s:socorg}.

\section{SoC's and FSO's as foundation for PSC environments}\label{s:found}
As we mentioned in Sect.~\ref{s:socorg},
PSC constitutes a ``communication channel'' (\emph{sensu\/} Boulding's model)
that is characterized by the following major services:
social translucence; referral services; and the empowering of a system's parts
and whole. In what follows we highlight how SoC and FSO support those services.

\subsection{Social translucence}\label{ss:st}
As already mentioned, social translucence is obtained in SoC's by identifying mutualistic
relationships and by making the involving parties aware of the existence of those relationships.
In order to better explain this concept we now introduce a number of definitions.

\begin{definition}[Social action]
Let $D$ and $R$ be two social systems.
Behaviors may take place in either system
as specified by behavior sets $B_D$ and $B_R$. Then the
following bijective function:
\begin{equation}
\sigma: B_D \to B_R
\label{e:act}
\end{equation}
maps behaviors in $B_D$ into corresponding behaviors in $B_R$.
\end{definition}

As an example, if $D$ and $R$ are respectively the animalia and plantae kingdoms,
then $\sigma$ shall map behaviors produced by animals (for instance,
respiration) into behaviors experienced by plants (for instance, production of
carbon dioxide); and if $D$ and $R$ are respectively the plantae and the animalia kingdoms,
then $\sigma$ shall map behaviors produced by plants (for instance,
photosynthesis) into behaviors experience by animals (for instance, production of
oxygen).

\begin{definition}
Let $S$ be a social system
in which behaviors may
take place as specified by behavior set $B_S$ . Then the following
function:
\begin{equation}
\varepsilon_S : B_S \to I_S
\label{e:eval}
\end{equation}
maps behaviors in $B_S$ into a semantic interpretation/evaluation of
the significance of those behaviors for $S$. We assume that said
interpretation may be associated at least with one of the
following three classes: positive, neutral, and negative,
meaning respectively that the mapped action is evaluated as
being beneficial, insignificant, or disadvantageous. Integers 1,
0, and -1 will be used to represent the above three classes
respectively.
\end{definition}

As an example, if $S$ is the animal kingdom, then a behavior
such as $b=$``production of oxygen'' would be considered as beneficial,
and therefore $\varepsilon_S(b) = 1$.

We can now define a mutualistic precondition:
\begin{definition}[Mutualistic precondition]
Let $D$ and $R$, $B_D$ and $B_R$, and $I_D$ and $I_R$
be defined as above. Then the following
conditions are called the mutualistic precondition (MP)
between $D$ and $R$:
\begin{equation}
\exists b\in B_D : \varepsilon_D(b)\ge0 \wedge \varepsilon_R(\sigma(b))>0 \label{e:mp1}
\end{equation}
\begin{equation}
\exists c\in B_R : \varepsilon_R(c)\ge0 \wedge \varepsilon_D(\sigma^{-1}(c))>0 \label{e:mp2}
\end{equation}
\end{definition}

The first formula, \eqref{e:mp1}, states that there exists a behavior in $B_D$ that is interpreted as
positive or neutral, though its occurrence
produces positive returns for $R$.
The second formula, \eqref{e:mp2}, expresses a dual condition: an action $c$ that is either neutral or positive
in $R$ translates in a beneficial action $\sigma^{-1}(c)$.

Animal respiration and plant photosyntesis are behaviors that fulfill MP.

\begin{definition}[Mutualistic relationship]
A mutualistic
relationship between two social systems $D$ and $R$
is defined as the social behavior occurring when $D$ and
$R$ enact individual behaviors that correspond to the mutualistic
preconditions \eqref{e:mp1} and \eqref{e:mp2}. When a mutualistic relation
exists between $D$ and $R$,
we shall write $D\ointclockwise R$.
\label{d:mr}
\end{definition}

What possibly happens in nature is that the positive returns
triggered by a certain behavior of $D$ stimulate in $R$ the
production of a dual behavior. The positive interpretation of the
latter in $D$ further stimulates the production of the former
actions, which consolidates the mutualistic
relationship between $D$ and $R$ (viz.  $D\ointclockwise R$).

\begin{definition}[Chain of mutualistic relationships]
Mutualistic relationships may take place also between three or more social systems;
a typical case that occurs in nature is that of chains of mutualistic
relationships:
\begin{equation}
\exists b\in S_0 : \left( \bigwedge_{i=0}^{t-1} \varepsilon_{S_i}(\sigma^i(b))\ge0 \right) \wedge
\varepsilon_{S_t}(\sigma^t(b))>0
\label{e:mp3}
\end{equation}
\begin{equation}
\exists c\in S_t : \left( \bigwedge_{i=1}^{t-1} \varepsilon_{S_{t-1-i}}(\sigma^{-i}(c))\ge0 \right) \wedge
\varepsilon_{S_0}(\sigma^{-t}(c))>0.
\label{e:mp4}
\end{equation}
\end{definition}

The first formula, \eqref{e:mp3}, states that there exists a chain of behaviors
that are interpreted as
positive or neutral by a corresponding chain of social systems; and that
the last behavior in the chain
produces positive returns for the social system at the end of the chain.
The second formula, \eqref{e:mp2}, expresses a dual condition: there exists
a chain of behaviors that ``moves'' from the end of the chain towards its
beginning; all those behaviors are positive or neutral, except at the beginning of the chain,
for which the behavior is beneficial.

When \eqref{e:mp3} and \eqref{e:mp4} both hold for a set of social systems $\mathbf{S}$, we
shall indicate this by means of symbol $\ointclockwise_\mathbf{S}$.
The same symbol shall be used also for any other form of mutualistic relationship
experienced by a set of social systems $\mathbf{S}$, including the one
introduced in Definition~\ref{d:mr}.

The above definitions and symbols allow us to construct
the following semi-formal definition of social translucence.
\begin{definition}[Social translucence]
Given a set of social systems $\mathbf{S}$ for which $\ointclockwise_\mathbf{S}$ holds,
social translucence is the property to making all involved social systems
aware that $\ointclockwise_\mathbf{S}$ holds.
\end{definition}

Achieving social translucence thus means that the SCE of a SoC makes use of the
\texttt{notify()} method of Fig.~\ref{f:SoC-1} to spread awareness
of the existence of a ``behaviour''
(in fact, a service request) that, once enacted, would result in a mutualistic relationship.
In case of global enrollment, the first \texttt{notify()} would trigger subsequent \texttt{notify()}'s
across the nodes of the involved SoC's until all the enrolled social systems are made aware
of the benefits of the social union expressed by the service request.

\subsection{Empowering the parts and the whole through e-referral}\label{ss:eref}
As mentioned before, a major advantage of FSO and the SON mechanism is given by
an e-referral approach that extends the scope of the one provided by SoC's.
When a request for services $a$ is enabled, the corresponding actors become a new temporary SoC whose
lifespan in limited to the duration of $a$. If enabling a request for services required
exceptions---in other words,
when the enrollment is global---the new temporary
SoC is made of agents from different communities. In
that case, the new community brings together nodes
from different layers of the FSO hierarchy. Because of
this, we call such a new SoC a social overlay network (SON).
The rigidity of existing organizations is thus
replaced by an agile and dynamic PSC environment that exploits
and promotes cooperation between PSC loci that represent
societal services at different scales.

The need for such enhanced cooperation may be proved by considering two
cases: a first one focusing on crisis management organizations, and a second
one related to healthcare.
The first case is one in which a single event with a global scope affects several
social organizations at the same time.

The second case is one in which a single event requires a complex, composite,
and coherent response from a number of social organizations.
In the rest of this section we shall briefly discuss those two cases
in Sect.~\ref{ss:crisis} and Sect.~\ref{ss:care}.

\subsubsection{Crisis management organizations}\label{ss:crisis}
A practical case where FSO's and their SON's may be particularly of use is that of
crisis management and recovery.
As observed in~\cite{DFSB14,DF15d}, disastrous events such as the Katrina hurricane~\cite{CARRI3}
disrupt several concentric ``social layers'' at the same time---typically local, regional, national,
and federal emergency response organizations. A major problem that was experienced during the
Katrina crisis was that those organizations may be loosely coordinated and non-cooperating.
Conflicting goals and conflicting actions;
multiple uncoordinated efforts that resulted in wasting resources and in
some cases masked each other out;
the inability to share promptly and dynamically the organizational assets
according to the experienced needs; 
and the inability to make use of spontaneous (that is, non-institutional) responders~\cite{DFSB14},
where some of the reasons that slowed down
and degraded considerably the effectiveness of the response to Katrina:
\begin{quote}
    [Responders] ``would have been able to do more if the tri-level system (city, state, federal)
of emergency response was able to
    \emph{effectively use, collaborate with, and coordinate the combined public and private efforts}''~\cite{RAND}.
\end{quote}
Similar delays and inefficiencies~\cite{Miskel08} were experienced also in other cases\footnote{%
	See for instance the case of Hurricane Andrew~\cite{Adair02}. Two eloquent quotes by Dr. Kate Hale,
	Dade County's emergency management director during Andrew's crisis, were
	``They keep saying we're going to get supplies. For God's sake, where are they?'';
	``\emph{Where in the hell is the cavalry on this one?}''.}.

A major technique advocated as a solution to organizational delays and efficiencies in disaster recovery
is so-called community resilience.
According to RAND~\cite{RAND}, community resilience is

\begin{quote}
A measure of the sustained ability of a community to utilize available resources to respond to, withstand, and recover from adverse 
situations. [\dots] Resilient communities withstand and recover from disasters. They also learn from past disasters to strengthen future recovery 
efforts.
\end{quote}

In~\cite{Paja}, an FSO for crisis management enabling community resilience was proposed.
By means of multi-agent simulations it was shown of two SoC's may share knowledge and resources
in the course of a crisis. The crisis in this case was modeled as a number of
houses catching fire. Timely intervention was necessary in order to contain the damage.
Said simulation proved that cooperation between fire fighters organizations and ``non-institutional''
responders (individuals in proximity) considerably reduces the amount of burned down houses.

\subsubsection{Inter-organizational cooperation in healthcare}\label{ss:care}
Regardless of its nature, any system is affected by its design assumptions. Our societies are no exception.
The emergence of sought properties such as economic and social welfare for all;
sustainability with respect to natural ecosystems;
and especially manageability and resilience, highly depends on the way social organizations are designed.
A typical case in point is given by traditional healthcare organizations.
A common assumption
characterizing those organizations is the adoption of a strict
client-server model. The major consequence of said assumption
is the lack of server-side orchestration of responses to the users' requests.
In other words,
it is the responsibility of the client to identify which server to bind to;
   it is the user that needs to know, e.g., which emergency service to invoke; which
   hospital to call first; which civil organization to refer to, and so on.
   Referral services \emph{do\/} exist, though they mainly cover
   very specific and simple cases (typically, the seamless transfer of patient information
   from a primary to a secondary practitioner~\cite{EReferrals}).
   Moreover,
   typically such services possess an incomplete view of the available resources.
Furthermore, to the best of our knowledge, none of the existing referral services provides a composite response to complex
   requests such that the action, knowledge, and assets
   of multiple servers are automatically or semi-automatically combined and orchestrated.
   Even electronic referral systems in use today are mostly limited~\cite{ShBe07} and only provide
   predefined services in specific domains\footnote{%
        Interesting examples of such systems include SHINE~\cite{SHINE} and SHINE OS$+$~\cite{SHINEOS+}.}.
   As a consequence,
   in the face of complex servicing requests calling for the joint action of multiple servers,
   the client is basically \emph{left on its own}. Societal organizations do not
   provide unitary responses nor assist the client in composing and managing them.
   Reasons for this may be found in lack of awareness and also in the ``convenient''
   shift of responsibility for failures
   from the server to the client\footnote{As observed by~\cite{ZenVerhulst},
        an example of said shift of responsibility may be found in car industry with respect to
        aviation industry. A matter for reflection is the fact that the shift of responsibility
        regrettably translates in an inferior safety culture.}.

Through the above considerations it becomes apparent that new and better
assumptions are called for by social organizations.
In particular, the increasing complexity of modern times require that societal
organizations assume responsibility for becoming the enablers
of collectively intelligent responses.
Those organization should function as a catalyst
of mutualistic cooperation among the role players at all levels,
from the citizens to the governing institutions. By means of the
organization, knowledge should flow among the players highlighting
needs, assets, requirements, and opportunities. The
organization should assist in the process of self-orchestrating a
response, making it easier for all parties involved to coordinate
themselves, exchange information, and take the right and
timely decisions.

Said new model has been studied
in the two papers~\cite{DeFPa15a,DeFPa15b}. There, we introduced two e-referral services
based on our FSO's.
The first service makes use of FSO exceptions and SON's to self-orchestrate
a composite response for the user. By means of multi-agent simulations we proved
that this considerably shortens the
average time an individual in need of care has to wait until s/he receives the necessary treatment,
increases the amount of treated patients,
and reduces the number of patients that could not timely receive their treatment.
A major conclusion is the empowerment of the individuals and of society as a whole.


The second service again uses simulated FSO's and their SON's to tackle the well-known problem
of falls identification. In order to improve the quality of
falls detection systems, a cloud of volunteers is used as an extra ``detection layer''
to verify whether alarms actually correspond to falls or are false positives.
Major conclusions are that FSO's dynamic hierarchical organization optimally
orchestrates all participating entities thus overcoming the stiffness of the traditional organizations.
Major returns include an
improvement of social costs and a
better use of the social resources (empowerment of the whole) as well as
a reduction of the average time to respond to identified falls (empowerment of the parts).

\section{Conclusions}\label{s:end}
We have recalled the major characteristics of SoC and FSO, and shown that they
correspond to the major requirements of pervasive social computing: social translucence,
e-referral, and the empowerment of a social system's parts and whole.
In particular, we have mapped social translucence with the social computing engine
of SoC's, and we have discussed FSO-based e-referral services in the crisis management
and healthcare management domains.
Thanks to the FSO organizational rules and FSO's nested compositional hierarchy,
the advertising of service requests extends beyond the originating locus. This makes it possible
to achieve ``inter-loci'' cooperation though the mechanism we called ``social overlay networks.''
We showed how
the ability to compose social overlay networks
constitute a mechanism to achieve complex e-referral services
and enable social cooperation between co-existing organizations.
By recalling results achieved through multi-agent simulation,
we suggested how FSO-based pervasive social computing replaces
the rigidity of existing organizations
with an agile and dynamic PSC environment that exploits
cooperation between PSC loci that represent societal services at different scales.
In our future work we hope to move from simulation to real-life experimentation
extending systems such as the FSO middleware of~\cite{DFSB13a,DF15c}.
A major challenge shall be to prove the resilience of those systems~\cite{DF15d,DF15b},
namely improving the quality of social organizations while preserving key
aspects of their identity.

%

\bibliographystyle{spmpsci}

\bibliography{/refs/thesis}
\end{document}